\documentstyle[manuscript,aps,psfig,cite]{revtex}

\begin{document}
\tighten
\draft
 \title{Perturbative treatment of intercenter coupling in Redfield
  theory} \author{Ulrich Kleinekath\"ofer, Ivan Kondov, and Michael
  Schreiber} \address{Institut f\"ur Physik, Technische Universit\"at, D-09107\ 
  Chemnitz, Germany} 
\date{\today} \maketitle
\begin{abstract}
  The quantum dynamics of coupled subsystems connected to a thermal bath is
  studied. In some of the earlier work the effect of intercenter coupling
  on the dissipative part was neglected. This is equivalent to a
  zeroth-order perturbative expansion of the damping term with respect to
  the intercenter coupling. It is shown numerically for two coupled
  harmonic oscillators that this treatment can lead to artifacts and a
  completely wrong description, for example, of a charge transfer processes
  even for very weak intercenter coupling.  Here we perform a first-order
  treatment and show that these artifacts disappear. In addition, we
  demonstrate that the thermodynamic equilibrium population is almost
  reached even for strong intercenter coupling strength.
\end{abstract}

\section{Introduction}
\label{sec:intro}
Quantum dynamics of complex molecules or molecules in a dissipative
environment has attracted a lot of attention during the last years. One
special kind of this problem is the electron transfer dynamics in or
between molecules especially in solution \cite{bixo99,newt91,barb96}. The
bath-related relaxation can be described in a variety of ways. Among others
these are the path integral methods \cite{weis99,makr98}, the semi-group
methods \cite{davi98b,kosl97,kohe97a}, and the reduced density matrix (RDM)
theory \cite{blum96}. The latter one has been especially successful in
Redfield's formulation \cite{redf57,redf65} and is the topic of the present
investigation.  As usual, the master equation for the RDM is derived from
the equation of the full system, i.e.\ relevant system plus bath, by
tracing out the bath degrees of freedom. The main limitations of Redfield
theory are the second-order perturbation treatment in system-bath coupling
and the neglect of memory effects (Markov approximation). In addition
Redfield suggested the use of the secular approximation.  In this
approximation it is assumed that every element of the RDM in eigenstate
representation (ER) is coupled only to those elements that oscillate at the
same frequency. In the present study we do not perform this additional
approximation which could distort the correct time evolution in transfer
problems\cite{barv99,barv00}.

To be rigorous in applying Redfield theory, the operators describing the
time evolution have to be expressed in ER of the relevant system as has
been done in the original papers \cite{redf57,redf65}. For electron
transfer this was performed in part of the literature (see for example
\cite{felt95,poll96,jean96,jean98}) while in another part of the literature
\cite{may92,may93,kueh94,fuch96,wolf95,wolf96} diabatic (local)
representations (DRs) have been used, which significantly reduces the
numerical effort in many cases.  In NMR literature
\cite{redf65,jeen95,cupe00} most people seem to use the ER while in quantum
optics most people use DRs \cite{cohe92,shor93}. Only recently the ER is
used in quantum optics \cite{cres92,mura95,zoub00}.  Here we focus on
electron transfer systems, but the conclusions should also be applicable to
problems in other areas.

While in ER the damping term is
evaluated exactly, in DR the influence of the
coupling between the local subsystems on dissipation is neglected. As a
consequence the relaxation terms do not lead to the
proper thermal equilibrium of the coupled system
\cite{harr85,davi98b,sega00a}.  Only the thermal equilibrium of each
separate subsystem is reached which can be quite different from the thermal
equilibrium of the coupled system. It will be shown here that even for a
very small intercenter coupling a completely wrong asymptotic value can be
obtained.

Although possibly leading to the wrong thermal equilibrium the local DR has
advantages.  For large problems it may be difficult to calculate the
eigenstates of the system. These are not necessary in the DR.  There one
only needs the eigenstates of the subsystems.  The quantum master equation
can be implemented more efficiently in DR in many cases
\cite{may92,wolf98,schr00,kond00}. Moreover, almost all physical and
chemical properties of transfer systems are expressed in the DR. For
example, to determine the transfer rate one often calculates the population of
the diabatic states and obtains the rate from their time evolution. To do
so one has to switch back and forth between DR and ER all the time if the
time evolution is determined in ER.

Using the semi-group methodology and a simple model of two fermion sites,
DR and ER have been compared already \cite{davi98b}. We are interested in a
more complicated system, i.e.\ a curve-crossing problem. The fact that we
use a different relaxation mechanism should effect the findings only very
little. Here we not only compare DR and ER but show how the relaxation term
in DR can be written more precisely for small intercenter coupling.

The paper is organized as follows. The next section gives an introduction
to the Redfield theory and, using the DR, presents a zeroth-order (DR0) and
a first-order (DR1) perturbation expansion in the intercenter coupling. In
the third section numerical examples are shown for two coupled harmonic
oscillators. The DR and ER results are compared to each other and also to
the improved local relaxation term derived here. The last section gives a
short summary.  Atomic units are used unless otherwise stated.

\section{Intercenter perturbation expansion within the Redfield equation}
\label{sec:eom}

In the RDM theory the full system is divided into a relevant system part
and a heat bath. Therefore the total Hamiltonian consists of three terms
-- the system part $H_{\rm S}$, the bath part $H_{\rm B}$, and the
system-bath interaction $H_{\rm SB}$:
\begin{equation}
H = H_{\rm S} + H_{\rm B} + H_{\rm SB}.
\label{eq:Hamiltonian}
\end{equation}
The RDM $\rho$ is obtained from the density matrix of the full system by
tracing out the degrees of freedom of the environment.  This reduction
together with a second-order perturbative treatment of $H_{\rm SB}$ and the
Markov approximation leads to the Redfield equation
\cite{redf57,redf65,blum96,may00}:
\begin{equation}
  \dot{\rho} = - i [H_{\rm S},\rho] + {\cal R}\rho 
={\cal L} \rho.
\label{eq:redfield}
\end{equation}
In this equation ${\cal R}$ denotes the Redfield tensor. If one assumes
bilinear system-bath coupling with system part $K$ and bath part $\Phi$
\begin{equation}
H_{\rm SB} = K \Phi
\label{eq:bath-coupling}
\end{equation}
one can take advantage of the following decomposition \cite{poll94,may00}:
\begin{equation}
\dot{\rho} = - i \left[ H_{\rm S},\rho \right]
+ \{
[ \Lambda\rho,K]+
[ K,\rho\Lambda^{\dagger}]
\}.
\label{eq:pf-form}
\end{equation}
Here $K$  and $\Lambda$
together hold the same information as the Redfield tensor ${\cal R}$.
The $\Lambda$ operator can be  written in the form
\begin{equation}
\Lambda=\int\limits_0^{\infty}d\tau
\langle\Phi(\tau)\Phi(0)\rangle K^{\rm{I}}(-\tau)
\label{eq:lambda}
\end{equation}
where $ K^{\rm{I}}(-\tau)=  e^{-i H t} K e^{i H t}$ is
the operator $K$ in the interaction representation.
Assuming a quantum bath consisting of harmonic oscillators
the time correlation function of the bath operator is given as
\cite{poll96}
\begin{eqnarray}
  C(\tau)=\langle\Phi(\tau)\Phi(0)\rangle=
\int\limits_0^{\infty} d\omega J(\omega) n(\omega)
 (e^{i \omega t}
+e^{\beta \omega} e^{-i \omega t})~.
\end{eqnarray}
Here $J(\omega)$ denotes the spectral density of the bath \cite{poll96},
$n(\omega)=(e^{\beta \omega}-1)^{-1}$ the Bose-Einstein distribution, and
$\beta=1/(k_{\rm B} T)$ the inverse temperature.

The  Hamiltonian $H_{\rm S}$ of the system we are interested in 
can be separated according to
\begin{eqnarray} 
H_{\rm S}=H_0+V
\end{eqnarray}
where $H_0$ is the sum of all uncoupled subsystem Hamiltonians $H_{0,n}$
\begin{eqnarray} 
H_0=\sum_n H_{0,n}
\end{eqnarray}
and $V$ the coupling among them which is assumed to be small.  Two
canonical bases can be constructed for such a Hamiltonian. One consists of
eigenfunctions of $H_0$.  It is often called a local basis because these
basis functions of the diabatic potential energy surfaces (PESs) are located
at specific subsystems (centers).  Latin indices such as $|n\rangle$ are
used below to denote these DR basis states. The other basis diagonalizes
the system Hamiltonian $H_{\rm S}$.  So it consists of eigenstates of
$H_{\rm S}$ and is called adiabatic basis. For these ER basis functions we
use Greek indices such as $|\nu\rangle$.  As discussed in the introduction
Redfield theory is defined in ER but for transfer problems DRs have some
conceptual and numerical advantages.

Here we first calculate the dissipation in the DR for small intercenter
coupling $V$. In this basis the matrix elements of $\Lambda$ are given by
\begin{eqnarray}
\label{lambda}
 \langle n| \Lambda|m\rangle = \int\limits_0^{\infty}  d\omega J(\omega)
n(\omega) \int\limits_0^{\infty} d\tau (e^{i \omega \tau}
+e^{\beta \omega} e^{-i \omega \tau}) \langle n|K^{\rm I}(-\tau)|m\rangle~.
\end{eqnarray}
To evaluate the matrix element of $K$ one has to use perturbation theory in
$V$ because the diabatic states $|n\rangle$ are not eigenstates of $H_{\rm
  S}$ but of $H_0$.  Some details of the determination of $\langle
n|\Lambda | m\rangle$ are given in the appendix.  Using the expression for
the correlation function in frequency space
\begin{eqnarray}
\label{eq:corr}
  C(\omega)=2 \pi [1+n(\omega)][J(\omega)-J(-\omega)]~,
\end{eqnarray}
and denoting the transition frequency between diabatic
states $|m\rangle$ and $|n\rangle$ by $\omega_{mn}$
the final result can be written as
\begin{eqnarray} \label{lambda-erg}
 \langle n|\Lambda |
m\rangle &=&\frac12 C(\omega_{mn})
 \langle n|K|m \rangle \nonumber \\
&& -\frac12 \sum_j \langle n |K|j \rangle \frac{\langle j|V|m \rangle}
{\omega_{jm}}
\left[ C(\omega_{mn})-  C(\omega_{jn})
\right] \nonumber \\
&& -\frac 12 \sum_i \langle i |K|m \rangle \frac{\langle n|V|i \rangle}
{\omega_{ni}}
\left[ C(\omega_{mn})-  C(\omega_{mi}) \right]~.
\end{eqnarray}
This first-order result DR1 can be split into a zeroth-order contribution
DR0 independent of $V$ and a first-order contribution proportional to $V$.
Taking the DR0 term
\begin{eqnarray} \label{lambda-erg0}
 \langle n|\Lambda |
m\rangle &=&\frac12 C(\omega_{mn})
 \langle n|K|m \rangle
\end{eqnarray}
only is equivalent to a complete neglect of the influence of the
intercenter coupling $V$ on dissipation.  This assumption has been used
earlier \cite{may92,may93,kueh94,fuch96,wolf95,wolf96} and is sometimes
called the {\it diabatic damping approximation} \cite{egor_pc00}.  In this
approximation only the states $|n\rangle$ and $|m\rangle$ contribute to the
matrix element $\langle n|\Lambda|m\rangle$. In DR1 all states contribute
to each of these matrix elements.  As a consequence the spectral
density of the bath is not only probed at the transitions of the uncoupled
subsystems as in DR0 but at many more frequencies.

The ER result for the matrix elements of $\Lambda$ can easily be deduced
from the DR result by replacing the diabatic states by adiabatic ones and
setting $V=0$ in Eq.~(\ref{lambda-erg}):
\begin{eqnarray}
 \label{lambda-erg-adi}
 \langle \nu|\Lambda |\mu{}\rangle &=&
\frac12 C(\omega_{\mu{}\nu})
 \langle \nu|K|\mu{}\rangle~.
\end{eqnarray}
This result is of course correct for arbitrary intercenter coupling
strength.

\section{Electron transfer in a two-center system}
\label{sec:etsystem}
In the following we direct our attention to electron transfer in an example
system consisting of two charge localization centers considered to be
excited electronic states. The PESs of the localization centers are assumed
to be harmonic and are sketched in Fig.~1.  For this example the
Hamiltonian of the uncoupled system is given by
\begin{eqnarray}
H_0=\sum_n \left[ U_n+ \left(a_n^{\dagger} a_n+\frac12\right) \omega_n \right]
\end{eqnarray}
and the coupling by
\begin{eqnarray}
V=\sum_{m,n} \sum_{M,N}(1-\delta_{mn})v_{mn}|mM\rangle \langle nN|~.
\end{eqnarray}
The first index in each vector denotes the diabatic PES while the second
labels the vibrational level.  $a_n$ and $a_n^{\dagger}$ are the boson
operators for the normal modes at center $n$ and $\omega_n$ are the
eigenfrequencies of the oscillators.  Bilinear system-bath coupling is
assumed and the system part is given by the coordinate operator $q$
\begin{equation}
K = q = \sum\limits_m \sum_{MN} \left( 2\omega_m {\cal M} \right)^{-1/2} 
\left( a_m^{\dagger} +a_m \right) |mM \rangle\langle mN|~
\label{eq:ka}
\end{equation}
The mass of the system is denoted by  ${\cal M}$.

In the local DR the system part of the system-bath coupling reads
\begin{equation}
\langle mM|K|nN\rangle=
\left( 2\omega_m {\cal M} \right)^{-1/2}\delta_{mn}
\left( \delta_{M+1,N}\sqrt{M+1}+\delta_{M-1,N}\sqrt{M}\right)~.
\label{eq:dia-ka}
\end{equation}
In the DR0 expansion (\ref{lambda-erg0}) the system can emit or absorb only
at intra-subsystem transition frequencies $\omega_{MN}$. The spectral
density of the bath $J(\omega)$ is effectively reduced to discrete values
$J(\omega)=\sum_m\gamma_m \delta(\omega-\omega_m)$.  The advantage of this
approach is the scaling behavior of the CPU time with the number ${\cal N}$
of basis functions which results from the simple structure of the $\Lambda$
matrix (\ref{lambda-erg0}).  As shown numerically \cite{kond00,schr00} it
scales like ${\cal N}^{2.3}$.  This is far better than the ${\cal N}^3$
scaling of the DR1 approximation (\ref{lambda-erg}).  In DR1 the
spectral density is probed at many more frequencies.  One needs the full
frequency dependence of $J(\omega)$ which we take to be of Ohmic form with
exponential cut-off
\begin{equation}
J(\omega)=\eta \Theta(\omega)\omega e^{-\omega/\omega_c}.
\label{eq:ohmic}
\end{equation}
Here $\Theta$ denotes the step function and $\omega_c$ the cut-off
frequency.  In this study all system oscillators have the same frequency
$\omega_1$ (see Table \ref{tab:parameters}) and the cut-off frequency
$\omega_c$ is set  equal to $\omega_1$. The normalization prefactor
$\eta$ is determined such that the spectral densities in DR and ER
coincide at  $\omega_1$.
Eq.~(\ref{eq:ohmic}) together with Eq.~(\ref{eq:corr}) yields the
full correlation function.

If the system Hamiltonian $H_{\rm S}$ is diagonalized and the resulting ER
basis is used to calculate the elements of the operators in
Eq.~(\ref{eq:pf-form}), there will be no longer any convenient structure in
$K$ or $\Lambda$, so that the full matrix-matrix multiplications are
inevitable.  For this reason the CPU time scales as ${\cal N}^3$, where
${\cal N}$ is the number of eigenstates of $H_{\rm S}$.  There appears to
be a minimal number ${\cal N}_0$ below which the diagonalization of $H_{\rm
  S}$ fails or the completeness relation for $|\nu\rangle$ is violated.
Nevertheless, the benefit of this choice is the exact treatment of the
intercenter coupling.  It is straightforward to obtain the matrices for
$\rho$ and $K$ (see for example Ref.\ \cite{felt95}).

 An initial wave packet at center $|n \rangle$ is prepared by a
$\delta$-pulse excitation from the ground state $|{\rm g}\rangle$ of the system
 \begin{equation}
\rho_{1M1N}(t=0)=
\langle 1M|{\rm g}0 \rangle\langle {\rm g}0|1N \rangle~.
\label{eq:rhoinit}
\end{equation}
 The pulse is
chosen such that mainly the fourth and fifth vibrational level of the
first (left) diabatic PES is populated. 
The motion of the
initial wave packet along the coordinate $q$ models the transfer
between the centers.  The parameters for our calculation are taken from
the work of K\"uhn et al. \cite{kueh94} and are shown in Table
\ref{tab:parameters}. Temperature is chosen as $T=295$ K and the reduced
mass of the system ${\cal M}$ is set to  20 proton masses.
The RDM is propagated in time and the occupation
probabilities for each localization center are calculated by means of the
partial trace:
\begin{equation}
P_m = \sum\limits_{M}\rho_{mMmM}.
\label{eq:populations}
\end{equation}
For the case of propagating in ER the RDM is transformed back to the DR in
order to apply Eq.~(\ref{eq:populations}).

In the following we compare the population dynamics in the two-center
electron transfer system using three different intercenter coupling
strengths $V$ and four different configurations of the two harmonic PESs.
The diabatic PESs and eigenenergies are shown in Fig.\ 1. Beginning our
analysis with the weak coupling case $v=v_{12}=v_{21}=0.1 \omega_1$ it is
expected that a perturbation expansion in $V$ yields almost exact results.
This is the reasoning why the DR0 term, which is easy to implement, has
been used in earlier work \cite{may92,may93,kueh94,fuch96,wolf95,wolf96}.

In configuration  (a) the eigenenergies of the two
diabatic PESs are in resonance. For example, the first vibrational
eigenenergy of the first center equals the third vibrational
eigenenergy of the second center. It is important to note that in this
configuration no vibrational level of the first center  is below the
crossing point of the two PESs. 
The calculations using ER and  DR0 as well as DR1 give almost identical
results, see Fig.~2a.  For long times DR0 deviates a tiny bit.  Redfield
theory in ER is known to give the correct long-time limit (up to the
Lamb shift).   

Configuration (b) differs from the first one by shifting the first PES up
by $\omega_1/2$. As shown in Fig.~2b the ER  and  DR1 results
again agree perfectly. On the other hand, the DR0 results are a little bit
off already at early times and the equilibrium value departs from the
correct value much more than in the first, on-resonance configuration.

Shifting the PESs further apart than in (a) yields configuration (c). The
energy levels are again on-resonance but this time two vibrational levels
of the first center are below the curve-crossing point, i.e.\ there is a
barrier for low-energy parts of the wave packet. As shown in Fig.\ 2c DR1
and the ER results agree perfectly once more. The DR0 results are terribly
off.  The long-time population of the first center which should vanish for
the present configuration stays finite.

If we increase the energy of the first PES by $\omega_1/2$ to obtain
configuration (d) DR0 fails again while DR1 gives correct results in
comparison to the ER, see Fig.~2d.

To understand the large difference between DR0 and DR1 we have a closer
look at the final result for the matrix elements of $\Lambda$, Eqs.\ 
(\ref{lambda-erg}) and (\ref{lambda-erg0}). The DR0 contribution
(\ref{lambda-erg0}) is independent of the intercenter coupling $V$. The
system part of the system-bath interaction $K$ allows only for relaxation
within each center.  So there is no mechanism in the dissipative part which
transfers population from one center to the other. This transfer has to be done
by the coherent part of the master equation. But the coherent part cannot
transfer components of the wave packet with energy below the crossing point
of the PESs. As tunneling is mainly suppressed, those components of the
wave packet cannot leave their center anymore although the corresponding
PES might be quite high in energy. This results in the failure of DR0 for
the configurations with barrier: Parts of the wave packet get trapped in
the two lowest levels of the left center.  From Eq.\ (\ref{lambda-erg}) one
can explain why in the on-resonance case the DR0 results are in better
agreement with the correct results. In this configuration some of the DR1
terms are very small and so the DR1 correction is smaller.

Now we discuss the medium coupling strength $v=0.5 \omega_1$ (see Fig.~3).
The results for configurations (a) and (b), i.e. without barrier, look
quite similar. In both cases the ER and DR1 results agree very well for
short and long times. At intermediate times there is a small difference. The
DR0 results  already deviate at short times and for long times there is
too much population in the left (higher) center. For configurations (c) and
(d), i.e.\ with barrier, again the ER and DR1 results coincide for
small and long times.  DR0 is off already after rather short times and the
long-time limit is again wrong.

For the strong coupling  $v=\omega_1$ (see Fig.~4) the behavior of the
results is quite similar to the medium coupling. For configurations (a) and
(b) the difference at intermediate times is a little larger, so is the
deviation of the long-time DR0 limit. For configurations (c)
and (d) with barrier there is also a discrepancy for DR1 already at short
times and the correct long-time limit is not reached exactly. But the
disagreement is surprisingly small for the strong coupling.  Overall DR1 still
looks quite reasonable while the DR0 results are completely off.

\section{Summary}

In addition to the approximations done in Redfield theory, i.e.\ 
second-order perturbation expansion in the system-bath coupling and Markov
approximation, we have applied perturbation theory in the intercenter
coupling.  It has been shown for two coupled harmonic surfaces that the
zeroth-order approximation DR0 which is equivalent to the {\it diabatic
  damping approximation} \cite{egor_pc00} can yield wrong population
dynamics even for very small intercenter coupling. These artifacts
disappear using the first-order theory DR1.

The scaling of DR1 is like ${\cal N}^3$ not as ${\cal
  N}^{2.3}$ for DR0. This is of course a serious drawback of DR1. For
configurations without barrier it seems to be possible to use DR0 for weak to
medium intercenter coupling. This of course depends on the accuracy
required especially for the long-time limit. In all other cases one should
either use the exact ER or DR1. Although the first-order results are not
exact for medium and strong intercenter coupling these calculations have at
least two advantages. First of all, one does not need to calculate the
eigenstates and energies of the full system Hamiltonian $H_{\rm S}$. For
small systems like two coupled harmonic surfaces using one reaction
coordinate this calculation is of course easy. But if one wants to study
larger systems like molecular wires \cite{davi98b,felt95} and/or multi-mode
models \cite{wolf95,wolf96,wolf98} this is no longer a trivial task. The
second advantage is related to the fact that in all transfer problems one
is mainly interested in properties which are defined in a local basis,
e.~g.\ the population in each subsystem in any moment in time. If one uses
the ER one has always to transform back to the DR in order to calculate
these properties. So for large-scale problems using a DR together with the
first-order perturbation in $V$ should be advantageous.

In a sense the present study is an extension of the investigation performed
by Davis et al.\ \cite{davi98b}. They compared ER and DR
for a two-site problem. Here we looked at a more general multilevel system
and also calculated the first-order perturbation. In their model
they do not have a reaction coordinate and therefore no
barrier. Their findings correspond more to cases (a) and (b) in
the previous section. Besides the agreement in the case of small
intercenter coupling they also found good agreement in the high-temperature
limit. Using our  model this statement could not be confirmed for
a general configuration, although there might be configurations where it is
true.  

In Ref.\ \cite{zoub00} the authors followed a strategy different from the
present work. They also studied two coupled harmonic oscillators modeling
two coupled microcavities, but only one cavity was coupled to the thermal
bath directly. This should not effect the questions studied here. With a
transformation to uncoupled oscillators they effectively reduced the
intercenter coupling to zero. The result \cite{zoub00} is then exact for
arbitrary  $V$.  The disadvantage of this strategy
is that it is not easy to extend to larger systems. The advantage of the
presently developed first-order expansion in $V$ is its general
applicability to problems of any size.

\acknowledgments
Useful discussions with V. May, W. Domcke, and D. Egorova are gratefully
acknowledged. We thank the DFG for financial support.

\appendix
\section*{}  
The purpose of this appendix is to show some more details for the
evaluation of $\langle n|\Lambda|m\rangle$. 
To calculate
\begin{eqnarray} \label{a1}
\langle n|K^{\rm I}(-t)|m\rangle &=&
\sum_{i,j} \langle n|e^{-i H t}|i \rangle \langle i|K|j\rangle
\langle j|e^{i H t}|m\rangle 
\end{eqnarray}
the  operator identity \cite{lair91}
\begin{eqnarray}
e^{-i (H_0 +V) t} =e^{-i H_0 t} 
\left( 1-i\int\limits_0^t dt' e^{i t' H_0} V e^{-i t'(H_0+V)} \right)~,
\end{eqnarray}
which can easily be proven by multiplying both sides by $e^{i H_0 t}$
and differentiating with respect to $t$, is used iteratively.
It yields 
\begin{eqnarray}
\label{trick}
\langle n|e^{-i H t}|i \rangle &=&
\langle n|e^{-i H_0 t} [1-i\int\limits_0^t dt' e^{i t' H_0} V e^{-i t' H_0}]
|i \rangle +{\cal O}(V^2)\nonumber \\
&=&e^{-iE_i t} \delta_{ni}-i e^{-i E_n t} \langle n|V|i\rangle
 \int\limits_0^t dt' e^{i (E_n-E_i) t'} +{\cal O}(V^2) \nonumber \\
 &=&e^{-iE_i t} \delta_{ni}-
\frac{\langle n|V|i\rangle}{E_n-E_i} (e^{-i E_i t}-e^{-iE_n t}) +{\cal O}(V^2)
\end{eqnarray}
assuming that $E_n \neq E_i$. Here and in the following we only give
the general expressions for the matrix elements. If a singularity can
appear due to coinciding frequencies the appropriate expression can be
obtained by taking the proper limit.

Thus the matrix element (\ref{a1}) is given by
\begin{eqnarray}
\label{tt}
\langle n|K^{\rm I}(-t)|m\rangle 
&=& e^{i \omega_{mn} t} \langle n|K|m\rangle \nonumber \\
&& -\sum_j \langle n|K|j \rangle \frac{\langle j|V|m\rangle}
{\omega_{jm}}(e^{i \omega_{mn} t}-e^{i \omega_{jn} t}) \nonumber \\
&& -\sum_i \langle i|K|m \rangle \frac{\langle n|V|i\rangle}
{\omega_{ni}}(e^{i \omega_{mi} t}-e^{i \omega_{mn} t}) +{\cal O}(V^2)
\end{eqnarray}
This result is inserted into Eq.\ (\ref{lambda}).
One has to evaluate integrals of the kind
\begin{eqnarray}
\int\limits_0^{\infty}dt e^{-\epsilon t} e^{-i \omega_{mn} t} 
=\frac{-i}{\omega-\omega_{nm}-i \epsilon}
\end{eqnarray}
which contain a convergence parameter $\epsilon$.
Using the well known identity
\begin{eqnarray}
   \lim_{\epsilon \to 0} \frac1{x \pm{}i\epsilon}
=\frac{P}{x} \mp \pi \delta(x)
\end{eqnarray}
one gets for the first term of the matrix element of $\Lambda$
\begin{eqnarray}
\langle n|\Lambda|m \rangle &=& 
 \frac{\pi}{1-e^{-\beta \omega_{mn}}}
[J(\omega_{mn})-J(-\omega_{mn})]
 \langle n|K|m \rangle  \nonumber + ({\rm Lamb}~{\rm shift}) + \ldots
\end{eqnarray}
The Lamb shift is the imaginary part of the matrix element
of $\Lambda$  and
leads to an energy shift in the quantum master equation.  This term is a
small correction \cite{rome89,geva00} and is neglected in Redfield theory.
The other terms of the  matrix elements are calculated in the same
fashion yielding
\begin{eqnarray} \label{lambda-erg2}
 \langle n|\Lambda |
m\rangle &=&\frac{\pi}{1-e^{-\beta \omega_{mn}}}
[J(\omega_{mn})-J(-\omega_{mn})]
 \langle n|K|m \rangle \nonumber \\
&& -\sum_j \langle n |K|j \rangle \frac{\langle j|V|m \rangle}
{\omega_{jm}}
\left\{
\frac{\pi}{1-e^{-\beta \omega_{mn}}} [J(\omega_{mn})-J(-\omega_{mn})]
\right. \nonumber \\ 
&& \left.-\frac{\pi}{1-e^{-\beta \omega_{jn}}} [J(\omega_{jn})-J(-\omega_{jn})]
\right\} \nonumber \\
&& -\sum_i \langle i |K|m \rangle \frac{\langle n|V|i \rangle}
{\omega_{ni}}
\left\{
\frac{\pi}{1-e^{-\beta \omega_{mn}}} [J(\omega_{mn})-J(-\omega_{mn})]
\right. \nonumber \\ && \left. 
-\frac{\pi}
{1-e^{-\beta \omega_{mi}}} [J(\omega_{mi})-J(-\omega_{mi})]\right\}
\end{eqnarray}

\bibliographystyle{prsty}


\begin{table}
\caption{Parameters used for the ground state oscillator and the two 
excited state oscillators.}
\label{tab:parameters}
\begin{tabular}{ccccc}
Center $|n\rangle$&  Configuration& $U_n$, eV& $Q_n$, ${\rm \AA{}}$&$\omega_n$, eV \\ \hline
$|{\rm g}\rangle$ &  &0.00 & 0.000 & 0.1 \\
$|1\rangle$ & &0.25 & 0.125 & 0.1 \\
$|2\rangle$ & a&0.05 & 0.238   & 0.1 \\
$|2\rangle$ & b&0.00  & 0.238   & 0.1 \\
$|2\rangle$ & c&0.05 & 0.363    & 0.1 \\
$|2\rangle$ & d&0.00  & 0.363    & 0.1
\end{tabular}
\end{table}


\begin{figure}
\psfig{figure=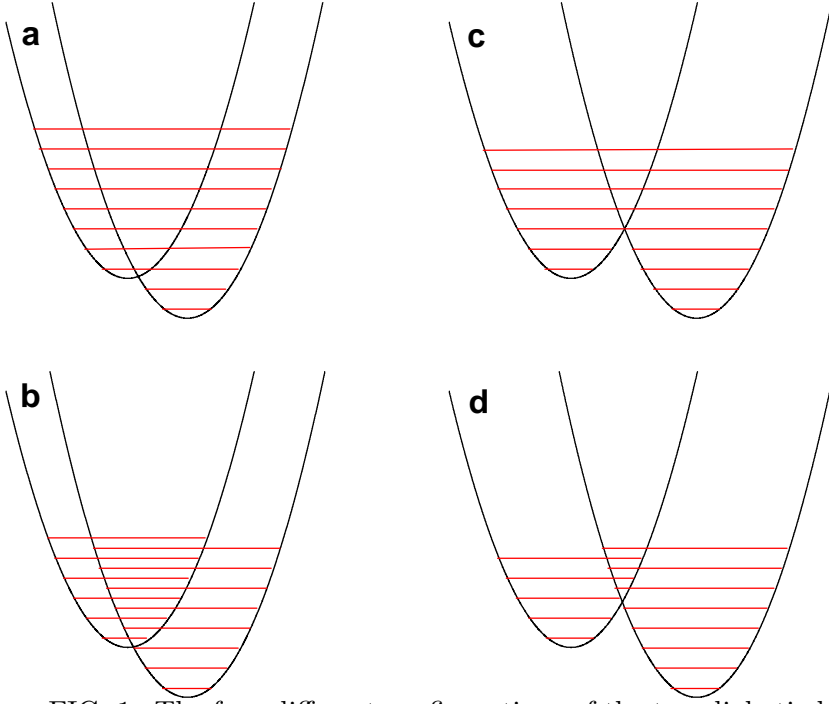,width=11cm}
\caption{The four different configurations of the two diabatic harmonic
  potentials $|1 \rangle$ and  $|2 \rangle$
 as discussed in the text. Also included in the figures are the
  energy levels. }
\end{figure}
\begin{figure}
\psfig{figure=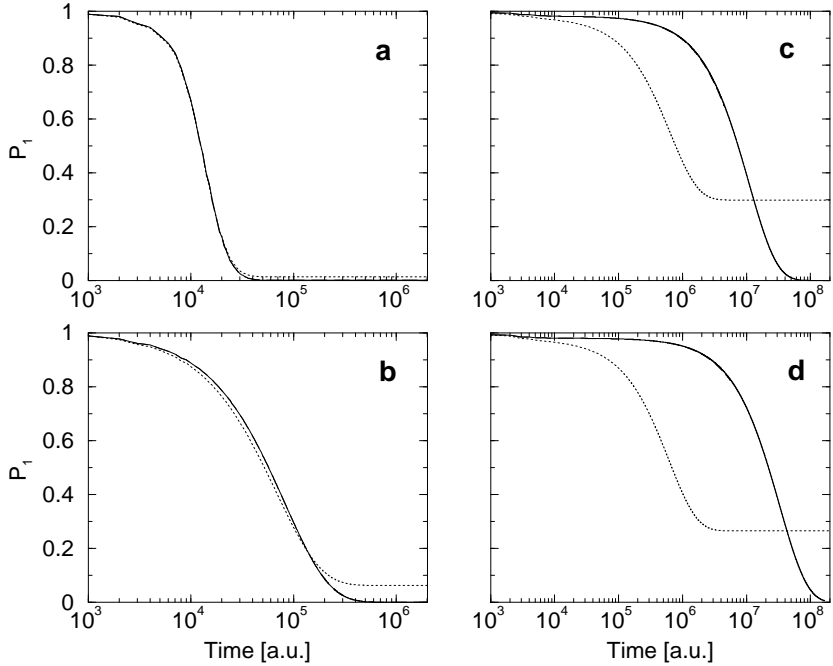,width=11cm}
\caption{Time evolution for small intercenter coupling and for the four
  different configurations. The results in ER are shown by
  the solid line while the results in diabatic basis are shown by dotted
  (zeroth-order) and dashed (first-order) lines. The results for ER and
DR1 are indistinguishable for small intercenter coupling.
 Note the logarithmic time
  scale.}
\end{figure}
\begin{figure}
\psfig{figure=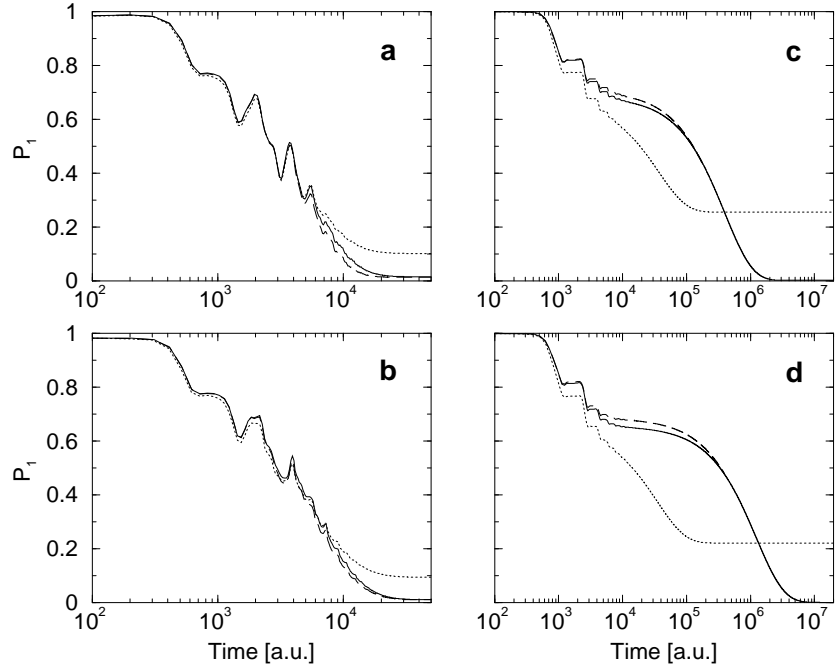 ,width=11cm}
\caption{Time evolution for medium intercenter coupling.}
\end{figure}
\begin{figure}
\psfig{figure=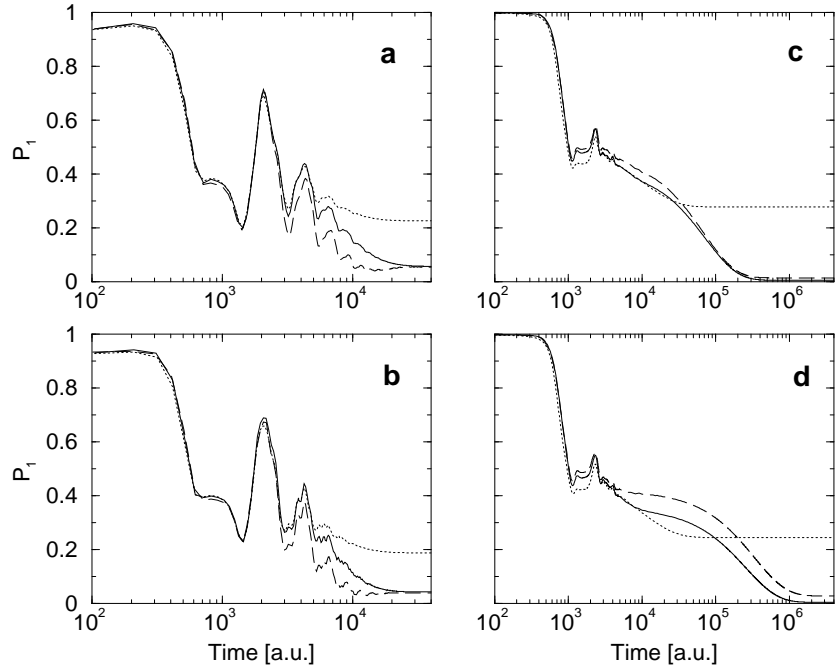,width=11cm}
\caption{Time evolution for strong intercenter coupling.}
\end{figure}

\end{document}